\documentclass[12pt]{article}
\usepackage{ragged2e}
\usepackage[left=18mm,top=26mm,right=18mm,bottom=18mm]{geometry}
\usepackage[affil-it]{authblk}
\usepackage{mathtools}
\usepackage{multirow}
\usepackage{lipsum}
\usepackage{graphicx,subfigure}
\usepackage{epstopdf}
\usepackage{amsmath, amssymb}
\usepackage{rotating}
\usepackage{morefloats}
\usepackage{comment}
\usepackage{caption}
\usepackage[square, numbers]{natbib}
\usepackage{booktabs}
\usepackage[english]{babel}
\usepackage{fancybox,color}
\usepackage{booktabs}
\usepackage[utf8]{inputenc}
\usepackage[english]{babel}

\newcommand\blfootnote[1]{%
  \begingroup
  \renewcommand\thefootnote{}\footnote{#1}%
  \addtocounter{footnote}{-1}%
  \endgroup
}

\begin{document}

\title{Mimetic Muscle Rehabilitation Analysis Using
Clustering of Low Dimensional 3D Kinect Data}

\author[1]{Sumit Kumar Vishwakarma }
\author[1]{Sanjeev Kumar}
\author[2]{Shrey Aggarwal}
\author[3]{Jan Mare\v s }

\date{}

\affil[1]{{\small Department of Mathematics, Indian Institute of Technology Roorkee-247667, India.}}
\affil[2]{{\small DIT Dehradun, India.}}
\affil[3]{{\small UCT Prague, Czech Republic.}}

% \author{Sumit Kumar Vishwakarma\textsuperscript{1}, Sanjeev Kumar\textsuperscript{1}, Shrey Aggarwal\textsuperscript{2}, Jan Mare\v s\textsuperscript{3}}
% \date{}

% \newcommand{\Addresses}{{% additional braces for segregating \footnotesize
% \bigskip
% \footnotesize
% \noindent \textsuperscript{1} Department of Mathematics, Indian Institute of Technology Roorkee-247667, India.
% \par\nopagebreak
% \noindent  \textit{e-mail:} \texttt{Sumit Kumar Vishwakarma: sumit\_kv@ma.iitr.ac.in} \\
% \noindent \textit{e-mail:} \texttt{Sanjeev Kumar: sanjeev.kumar@ma.iit.ac.in}\\
% \textsuperscript{2} DIT Dehradun, India.
% \par\nopagebreak
% \noindent  \textit{e-mail:} \texttt{Shrey Aggarwal : shreyagg2202@gmail.com}\\
% \textsuperscript{3} UCT Prague, Czech Republic. \par\nopagebreak
% \noindent  \textit{e-mail:} \texttt{Jan Mare\v s : Jan.Mares@vscht.cz}

% %\noindent \textsuperscript{*}Corresponding author.

% }}
\maketitle

\begin{abstract}
Facial nerve paresis is a severe complication
that arises post-head and neck surgery; This results in articulation problems, facial asymmetry, and severe problems in non-verbal communication. To overcome the side effects of post-surgery facial paralysis, rehabilitation requires which last for several weeks. This paper discusses an unsupervised approach to rehabilitating patients who have temporary facial paralysis due to damage in mimetic muscles. The work aims to make the rehabilitation process objective compared to the current subjective approach, such as House-Brackmann (HB) scale. Also, the approach  will assist clinicians by reducing their workload in assessing the improvement during rehabilitation. This paper focuses on the clustering approach to monitor the rehabilitation process. We compare the results obtained from different clustering algorithms on various forms of the same data set, namely dynamic form, data expressed as functional data using B-spline basis expansion, and by finding the functional principal components of the functional data. The study contains data set of 85 distinct patients with 120 measurements obtained using a Kinect stereo-vision camera. The method distinguish effectively between patients with the least and greatest degree of facial paralysis, however patients with adjacent degrees of paralysis provide some challenges. In addition, we compared the cluster results to the HB scale outputs.\\


\blfootnote{Email addresses: sumit\_kv@ma.iitr.ac.in (Sumit Kumar Vishwakarma),  sanjeev.kumar@ma.iit.ac.in (Sanjeev Kumar), shreyagg2202@gmail.com (Shrey Aggarwal), and Jan.Mares@vscht.cz (Jan Mare\v s) } 

\end{abstract}

\section{Introduction} %% {{{

Rapid growth in the development of modern sensors, storage devices, and technology opens up a new field of study driven by data. Also, with time this field has gained a certain level of maturity. The vast amount of data collected by modern sensors play an essential role in solving problem-related to medicine, diagnosis, and economics. 

\par This work is followed by the recent work given in \cite{b1} and \cite{b2}, which deals with the rehabilitation process after head and neck surgery causes temporary facial paralysis. Any patient who has undergone a surgical procedure on the head and neck has a probability of having facial nerve paralysis as an aftereffect, which causes due to damage in mimetic muscles linked by the facial nerve (cranial nerve VII).

\par Facial paralysis, commonly known as Bell's palsy, can be caused by various factors, including viral infections and post-surgical consequences. Trauma or injury to the face, head, or neck can induce paralysis by damaging the facial nerve. Tumors that grow on or near the facial nerve can compress or damage the nerve, leading to facial paralysis. Some persons may be born with anomalies or abnormalities of the facial nerve. Facial paresis is accompanied with a variety of physical issues, including trouble with face expressions, difficulty with eye closure, difficulty with speech and eating, hearing issues, migraines, and severe facial muscle atrophy \cite{b3}. The cases of peripheral facial palsy are sporadic in the medical field, ranging from 0.02 - 0.03 \% yearly. Since the problem involves facial expression, it also significantly impacts a person's social life. It can affect employment challenges, non-verbal communication, and social interactions, leading to emotional distress and social isolation \cite{b4}.

\par Evaluation of facial paralysis (FP) and quantitative grading of facial asymmetry is required to measure the condition's severity and track its improvement or advancement. Consequently, an accurate quantitative grading system that is user-friendly, affordable, and has minimum interobserver variability is required \cite{b5}. In \cite{b5}, the authors develops a comprehensive automated approach for quantifying and grading facial paralysis (FP). Recent advances  in automatic medical diagnosis using face images based on computer vision techniques provides unobtrusive objective information on a patient’s condition. The  work in \cite{b6} highlights the efforts that are being made to create robust systems suited for healthcare applications, addressing issues such as real-time evaluation. It deals with the most relevant and innovative solutions in facial images analysis.

\par The evaluation of facial nerve paralysis in patients enables clinicians to learn the severity of the condition. Also, it makes the clinician very efficient in comparing and discussing the severity with colleagues and addressing the rehabilitation. The House-Brackmann (HB) classification scale has been one of the most standardized for evaluating facial nerve paralysis, but it has some limitations concerning precision and inter-observer reliability. Other grading scales in literature are Botman and Jongkees Scale, May Scale, Pietersen Scale, and Smith Scale \cite{b7}.
\par Currently, the House-Brackmann (HB) classification (Table. \ref{Tab1}) scale is most widely used to evaluate facial nerve dysfunction by clinicians. The HB scale classifies the patients into six groups depending on the facial movement of the patients.
\par Patients with HB grade I have entirely regular facial movements with no synkinesis. If the patient is in HB grade II, it has only slight asymmetry during facial movement synkinesis. HB grade III patients have an asymmetry with some forehead movement. Patients in HB grade IV also have an asymmetry but no forehead movement and disfiguring synkinesis. If there is only slight movement, no forehead movement, and incomplete eye closure, patients are in HB V grade. In the absence of any movement, the patients are said to be in HB grade VI, and these patients are having total paralysis.
\begin{table}
\centering
\caption{House Brackmann (HB) Scale.}
\begin{tabular}{p{1.5cm}p{6cm}p{9.5cm}}
\hline
 Grade&Description& Characteristic \\
 \hline
I &Normal function  &Normal facial function in all nerve branches\\
II   & Mild dyfunction & Gross: slight weakness on close inspection; may have very slight synkinesis.\\
    & & At rest: normal symmetry and tone.\\
    & & Motion Forehead: moderate to good function.\\
    & & Eye: complete closure with minimum effort.\\
    & & Mouth: slight asymmetry\\
III&Moderate dyfunction & Gross: obvious but not disfiguring difference between two sides; noticeable synkinesis \\
 & & At rest: normal asymmetry and tone.\\
 & & Motion Forehead: slight to moderate movement.\\
 & & Eye: complete closure with effort.\\
 & & Mouth: slightly weak with maximum effort \\
IV &Moderately severe dyfuntion  & Gross: obvious weakness and/or disfiguring asymmetry.\\
 & & At rest: normal asymmetry and tone.\\
 & & Motion Forehead: none.\\
 & & Eye: incomplete closure.\\
 & & Mouth: asymmetric with maximum effort \\
V& Severe dyfunction& Gross: only barely perceptible motion.\\
& & At rest: asymmetry.\\
& & Motion—Forehead: none.\\
& & Eye: incomplete closure.\\
& & Mouth: slight movement. \\
VI& Total paralysis&No movement.\\
 \hline
\end{tabular}
\label{Tab1}
\end{table}

\par The study in \cite{b1} involves a three-step statistical analysis of post-surgical rehabilitation. The first step entails finding the appropriate indicator that expresses the facial nerve recovery, which involves symmetry, intensity, and speed curves. In the second step, the authors applied Functional Logistic Regression (FLR) to find a set of variables, referred to as the Health score, based on facial nerve indicators and the HB grade assigned by the clinician. The final step involves the classification of patients based on the Health score variables and HB grade through the application of ordinal logistic regression.
\par Several supervised learning methods are studied 
in \cite{b2}. The study involves parametric, non-parametric, and
neural network methods to classify the patients based on
Health score variables, as in \cite{b1}. The overall Comparison of the different supervised learning indicates that the parametric approach (Ordinal Logistic Regression) gives the best results in terms of accurate grading of facial paralysis. In both the studies discussed above, the HB grade is a significant factor in determining the Health score variable for the classification, which involves the subjective opinion of the physicians
\par In this paper we aims to look for an alternative and more objective method of grading facial paresis in patients, which would aid medical personnel in monitoring the rehabilitation process and implementing the necessary further medical procedures. This study uses complete unsupervised learning to determine the facial grade of patients based on the facial nerve indicator curve and compare different algorithms based on different forms of the data set. We performed clustering for dynamic data, functional data expressed using basis expansion, and principal component analysis of functional data (FPCs). This approach may also help in self-monitor the rehabilitation process and track the progress.
\par The paper is organized as follows. Section \ref{FDA} briefly discussed functional data and its components, such as basis expansion and FPCA. Section \ref{method}, discuss the data acquisition, preprocessing and various clustering algorithms. We implement clustering algorithms on the data set and compare the results in Section \ref{result}. In Section \ref{Discussion} we give a brief discussion, and future scope of the work concludes the paper in section \ref{future work} .
%%%%%%%%%%%%%%%%%%%%%%%%%%%%%%%%%%%%%%%
%%%%%%%%%%%%%%%%%%%%%%%%%%%%%%%%%%%%%%%%%%%%%%
\section{Functional Data analysis}\label{FDA}

\par Medical, biological, economic, and environmental problems increasingly incorporate inter-disciplinary approaches that use signal processing, imaging technologies, statistical methodology, and modern machine learning techniques. Functional data are a sort of consistently recorded data collection, and its statistical analysis is known as functional data analysis (FDA) \cite{b8}.
\par
\begin{figure}[!t]
\centerline{\includegraphics[width=\columnwidth]{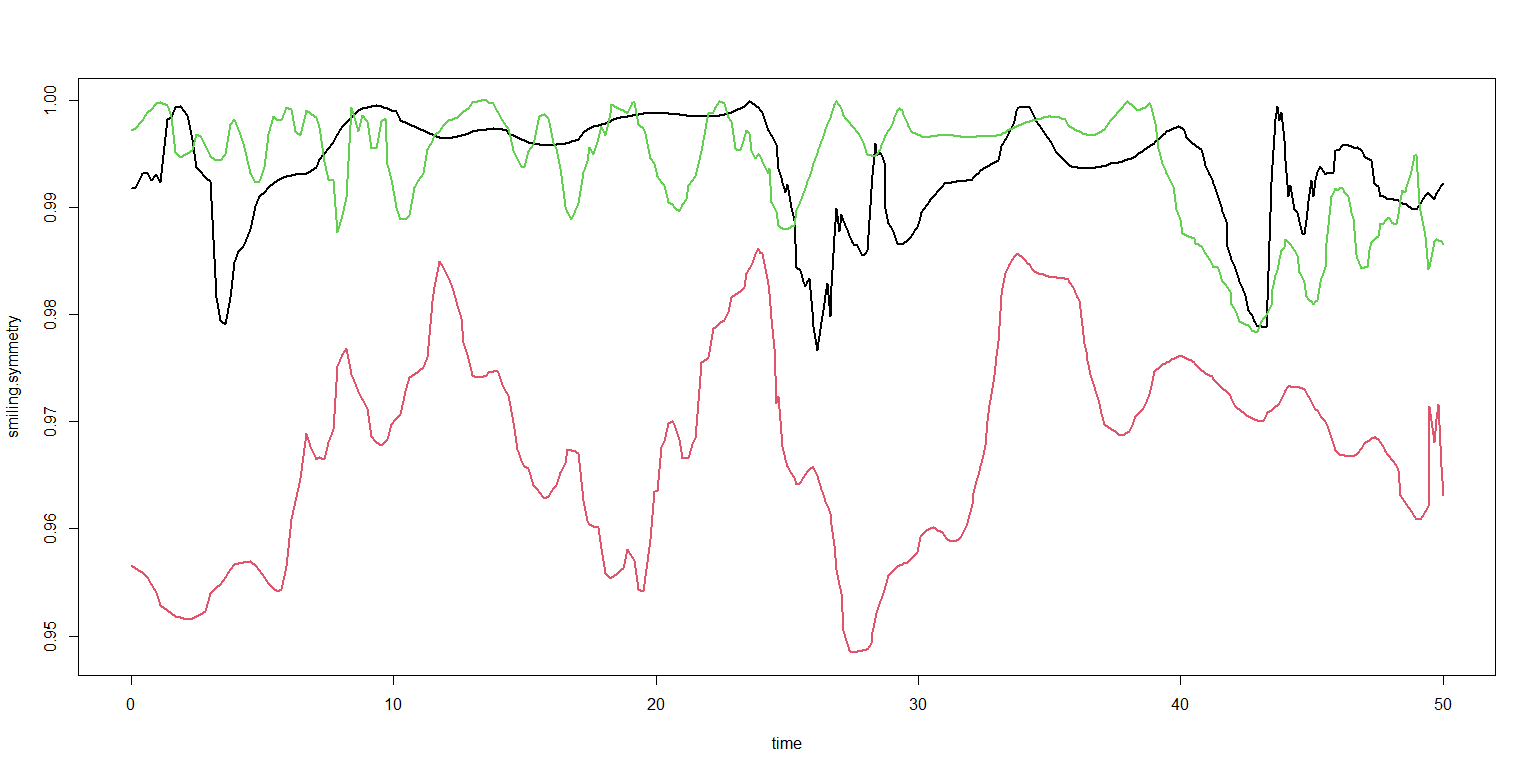}}
\caption{Example of functional data (Shows symmetry indicator of three patients).}
\label{fig1}
\end{figure}

 Functional data can be formally represented as a curve $(t,X)$, where $t$ represents the time domain and $X$ represents a  functional entry (Fig. \ref{fig1} shows three functional data).
FDA executes statistical analysis on the smooth curve 
\begin{equation*} 
X = X(t) \qquad t \in [T_1, T_2], 
\end{equation*}
such that values of $X(t)$ exist for all $t$.

%%%%%%%%%%%%%%%%%%%%%%%%%%%%%%%%%%%
\subsection{Functional data} \label{FD}
Functional data $\{X_1(t), X_2(t), X_3(t),...,X_N(t)\}$ can be thought of as a set of i.i.d samples of random variables on an interval $[T_1, T_2]$, with values in an infinite dimensional space. It is widely known that this type of data is regarded as a continuous-time stochastic process in a Hilbert space $H$. It is referred to as single variate functional data if the elements of $H$ are real valued functions. It is multivariate functional data if the elements of $H$ are $\mathbb{R}^p$-valued functions, where $p \geq 2$ \cite{b9}.
\par Functional data are supposed to belong to an infinite dimensional space, but in practice, we only have the discrete observation $X_{ij}$ for each sample curve $X_i(t)$ at time points $t_{ij}$; $j=1,2,3...,m_i$. A general assumption regarding the sample path is that it belongs to a finite dimensional space and may be spanned by some basis function to approximate the infinite-dimensional functional data into finite dimension. Therefore, $X_i$ can also be approximated as:
\begin{equation*} \label{eq:1} X_i=a_{i1}\beta_1(t)+a_{i2}\beta_2(t)+a_{i3}\beta_3(t)+...+a_{iK}\beta_K(t)
\tag{1}
\end{equation*}

where $\beta = \left\{ \beta_1., \beta_2,..., \beta_K \right\}$ is a basis, $K\in \mathbb{N}$, $a_{ik} \in \mathbb{R}$ are the coefficients of the basis expansion that can be calculated from discrete observations using a numerical method. Examples of well known basis are B-spline, Fourier basis, polynomial basis and many more. 
Thus,
\begin{equation*}
    X_{ij} = X_i(t_{ij}) = \sum_{k=1}^{K}{a_{ik}\beta_k(t_{ij})}  \quad j=1,2,3,...,m_i
\end{equation*}
\par In the realm of computational technology, it is a common occurrence for machines to collect data that contain inaccuracies or errors. This phenomenon can occur due to a variety of factors, including limitations in the data collection process, or inaccuracies in the sensors or instruments used to gather the information. Thus, if the sample curve are observed with error then,
\begin{equation*}
    X_{ij} = X_i(t_{ij}) + e_{ij}  \hspace{0.5cm} j=1,2,3,...,m_i
\end{equation*}
where $e_{ij}$ is the random noise follows Gaussian distribution with mean zero and variance $\sigma ^2$.

%%%%%%%%%%%%%%%%%%%%%%%%%%

\subsection{Functional principal component analysis}\label{FPCA}
Principal component analysis (PCA) is a well known and accepted tool in machine learning to reduce the dimension of multivariate data; FPCA is analogous to the PCA for functional data. Consider the space of the square integrable function, $L_2$ on $[T_1, T_2]$. The primary step in determining the functional principal component for functional data $X$ is to assumes that it is an $L_2$ continuous stochastic process. Therefore, for all $t \in [T_1, T_2]$, we have :
\begin{equation*}\label{eq:2}
    \lim_{h \to 0}\mathbb{E}\left[ |X(t+h)-X(t)|^2\right] = 0 
    \tag{2}
\end{equation*}
where $\mathbb{E}$ is expectation operator.\\
Let $\mu = \mu(t)= \mathbb{E}(X(t))$  signify the mean of the  function $X$. The covariance operator $\nu$ of $X$ is given by:
\begin{equation*}
    \nu:  L_2([T_1, T_2]) \to  L_2([T_1, T_2])
\end{equation*}  

\begin{equation*} \label{eq:3}
      f \xmapsto{\nu} \nu f = \int_{T_1}^{T_2}{V(.,t)f(t)dt}
\tag{3}
\end{equation*}
is an integral operator with kernel $V$ defined by:
\begin{equation*}
    V(s,t) = \mathbb{E}[(X(s)-\mu (s))(X(t)-\mu(t))], \hspace{0.5cm}s,t \in [T_1, T_2]
\end{equation*}
The mean and covariance function are continuous, and the covariance operator $\nu$ is a Hilbert-Schmidt operator.
\par The spectral analysis of $\nu$ provides a countable set of positive eigenvalues $\{\lambda_j\}_{j \geq 1}$  with $\lambda_1 \geq \lambda_2 \geq \dots $, associated to an orthonormal basis of eigenfunctions $\{f_j\}_{j \geq 1}$:
\begin{equation*}
    \nu f_j = \lambda_jf_j,
\end{equation*}

\begin{align*}
\int_{T_1}^{T_2} {f_j(t) f_{j'}(t)} dt = 
\begin{cases}
1 & \text{if $j = j'$} \\
0 & \text{otherwise}.
\end{cases}
\end{align*}

The principal components $\{C_j\}_{j \geq 1}$ of $X$ are random variable defines as the projection of $X$ on the eigenfunctions of $\nu$ :
\begin{equation*}\label{eq:4}
    C_j = \int_{T_1}^{T_2} (X(t)-\mu(t))f_j(t)dt.
\tag{4}
\end{equation*}
The principle components  $\{C_j\}_{j \geq 1}$ are zero mean uncorrelated random variable with variance $\lambda_j$, $j \geq 1$. \cite{b10}

\par From \cite{b11}, we see using the results in equations \ref{eq:3} and \ref{eq:4}, with the Karhunen-Loeve expansion \cite{b12}, the following result holds: 

\begin{equation*}
    X(t) = \mu(t) + \sum_{j \geq 1}C_jf_j(t), \hspace{0.5cm} t\in [T_1, T_2]
\end{equation*}
Alternatively, to utilize the classical ML tools of multivariate data analysis, we can truncate the above equation at first $q$ terms as below:
\begin{equation*}
    X(t) = \mu(t) + \sum_{j=1}^{q}C_jf_j(t), \hspace{0.5cm} t\in [T_1, T_2]
\end{equation*}
More details about functional data and FDA with statistical overview can be found in \cite{b11,b13}.

%%%%%%%%%%%%%%%%%%%%%%%%%%%%%%%%%%%%%%%%%%%%%%%%
%%%%%%%%%%%%%%%%%%%%%%%%%%%%%%%%%%%%%%%%%%%%%%%%%
\section{Proposed Methodology} \label{method}
The study involved 120 measurements of 85 patients who had gone head and neck surgery with certain probability of postoperative facial nerve dysfunction or the patients already had facial nerve palsy.

\subsection{Data Acquisition} \label{Data}
The data were collected by a mobile robotic system, which operates in static mode (also in dynamic mode) to collect the data from patients before and after surgery over a defined schedule. During the process, the patients were asked to perform several facial movements (Exercise mentioned in Table. \ref{Tab2}), which involves the mimetic muscles, while sitting in front of the Kinect sensor camera. At the same time, the clinician also evaluates the patients; based on the mimetic moment of the patients, the clinicians grade them as per the HB scale. This method of labeling the patients by physicians is not based on some robust mechanism but instead based on their subject expertise and experience, making the whole process highly subjective.

\begin{table}
\centering
\caption{Exercise to be performed by patients.}
\begin{tabular}{p{4cm}p{8cm}}
 \hline
 Exercise Name & Description \\
 \hline
 Raising   & Raise your eyebrow\\
 Frowning &  Frown\\
 Closing & Close your eyes tightly\\
 Smilig  & Smile at camera \\
 Baring &  Bare your teeth\\
 Pursing & Purse your lips\\
 Blowing & Blow out your cheeks \\
 closing  and Baring & Close your eyes tightly and bare the teeth\\
 Raising and Pursing & Raise your eyebrows and purse the lips\\
 \hline
\end{tabular}
\label{Tab2}
\end{table}

The Kinect sensor focuses on 21 POI (Point of interest)  of the patient's face (see Table. \ref{Tab3}). It measured 3D face data for each mentioned exercise in Table. \ref{Tab2} and stored it in a text-based file for further offline analysis.
\par The Kinect sensors collect data about patients in a  multivariate dynamic format. For each patients, we have $9$ exercise which has three dimensional measurement for each of $21$ facial point which are POI i.e., We have $567$ $(21\times 3 \times 9)$ time curves for each patient (measurement). High-dimensional data is accessible for each patient, but we have just 120 samples, which poses a substantial risk of model overfitting. Reducing the dimension of data and gaining access to only the qualities that are relevant to face nerve recovery can serve the aim admirably. Therefore, it is necessary to do appropriate preprocessing on the data set that explains face nerve recovery-specific factors.
 
\begin{table}
\centering
\caption{ Points of interest (internal index number)}
\begin{tabular}{ p{1cm}p{4cm}p{1cm}p{6cm}}
  \hline
 POI & Position & POI & Position \\
 \hline
0& left eye, bottom &11& left eyebrow, centre\\
1&left eye, top&12&nose tip\\
2&left eye, inner corner&13&mouth lower lip, central-bottom\\
3&left eye, outer corner&14&mouth, left corner\\
4&left eyebrow, inner &15& mouth, right corner\\
5 & left eyebrow, centre&16&mouth upper lip, centre-top\\
6&right eye, bottom&17&chin, centre\\
7&right eye, top&18&forhead, centre\\
8&right eye, inner&19&left cheek, centre\\
9&right eye, outer&20&right cheek, centre\\
10&right eyebrow, inner &&\\
 \hline
\end{tabular}
\label{Tab3}
\end{table}

\subsection{Data Preprocessing} \label{preprocessing}

\par The patients are being recorded by the 3D Kinect sensor camera while performing the suggested exercises. In order to extract relevant information regarding facial recovery, there is a need to identify specific curves that indicates the most significant status of facial paresis. For each exercise, one distance curve is identified that exhibits the most significant changes, for example, the distance between mouth corners for Smiling or the distance between the eyebrow and inner eye corner for Raising. Since, patients perform the exercise at different speeds and at different times; therefore, there is a need to align the temporal curves which express the facial exercises most significantly.
Let $P_i(t)$ denotes a curve which represents a particular exercise for measurement $i$, $P_i(t)$ is transformed using a time-warping function $w_i$ to $\tilde P_i(\tilde t)$. i.e.,
\begin{align*}
\tilde P_i(\tilde t) = P_i(w_i (\tilde t))
\end{align*}
where $\tilde t = w_i^{-1}(t)$, such that $\tilde P_i$ is aligned for all measurements $i$. The landmark registration technique, which aligns landmark points via piecewise cubic interpolation, is used to construct the warping function $w_i$.

\par To restrict the volume of data, it is necessary to utilize particular indicators that assess the symmetry (for comparing the left side of the face with the right side of the face), intensity (for the range of motion during exercise), and speed (which detect how fast or slow the exercise is performed) for each measurement. This indicator can be obtained for each exercise and denoted, for example, as smiling.symmetry, smiling.intensity, smiling.speed. The details of the computation of these indicators can be found in \cite{b1}. 

\par To verify that various indicator curves represent the level of sickness of the patients, Spearman's correlation between HB grade and health score (a numerical value that represents the healthiness of patients based on the indicator curve) is computed. It is found that smiling.symmetry and Baring.symmetry has a strong correlation (0.5 \& 0.47 resp.) with HB grade (cf. Table 8 in \cite{b1}). This implies that symmetries of exercises with the mouth represent the best correlation with facial nerve dysfunction, both for the clinician and the Kinect sensor. Therefore, we utilize this smiling.symmetry indicator as our feature for clustering the patients into different grades (level) of facial paralysis. 

\begin{table}
\centering
\caption{Frequency of HB grades.}
\begin{tabular}{ p{4cm}p{1cm}p{1cm}p{1cm}p{1cm}p{1cm}p{1cm}p{1cm}}
 \hline
 & HB1& HB2 &HB3& HB4& HB5& HB6& Total \\
 \hline
No. of patients&57&20&18&05&05&15&120\\
\hline
\end{tabular}
\label{Tab4}
\end{table}

\par Our data set is labelled (Table. \ref{Tab4}) by clinician into six categories, HB1-HB6, and the frequency of the HB4 and HB5 patient classes is relatively low, as these two classifications are rarely utilised. To achieve parity, we can reclassify the HB4 class as HB3 and the HB5 class as HB6, as they are rather comparable in terms of mimetic movement. Therefore, we reclassify our data set using the modified HB grades (Table. \ref{Tab5}). It will also assist us in comparing our clustering results to clinical labelling.

\begin{table}
\centering
\caption{Adjusted frequency of HB grades.}
\begin{tabular}{ p{4cm}p{2cm}p{2cm}p{2cm}p{2cm}p{2cm}}
  \hline
 & HB1 & HB2&HB3& HB6& Total \\
 \hline
No. of patients&57&20&23&20&120\\
\hline
\end{tabular}
\label{Tab5}
\end{table}
%%%%%%%%%%%%%%%%%%%%%%%%%%
\subsection{Clustering approach} \label{clustering}
 In machine learning, the clustering of functional data is termed unsupervised learning. The goal of clustering is to identify groupings of data that are similar within the group but distinct between the groups. The Clustering algorithm, such as $k$-means and hierarchical clustering, can be extended from vector-valued data to functional data, but with additional consideration, such as discrete approximation of distance measure and dimension reduction of infinite dimensional functional data. Usually clustering of functional data is very challenging.
 
\par Data labeling is always an expensive job to do, data related to medicine requires even more effort to label, and there is a high chance that the labeling is subjective. To overcome this and keep the rehabilitation process objective in nature, unsupervised learning plays an important role in this aspect.

\par For example, $k$-means functional clustering method seeks to identify a collection of $k$ cluster centers, 
\begin{equation*}
   \{\mu^c; \quad c=1,2,3...,k\},
\end{equation*}
 for a given sample of functional data
 \begin{equation*}
    \{X_i ; \quad i=1,2,3...,n\},
\end{equation*}
 This is achieved by minimizing the sum of squared distances between $\{X_i\}$ and the cluster centre associated with their cluster labels $\{C_i\}$.
That is, the $n$ functional data $\{X_i\}$ are partitioned into $k$ groups such that: 
\begin{equation*}
    \frac{1}{n}\sum_{i=1}^{n} d^2(X_i, \mu_{n}^c),
\end{equation*}
is minimized over all possible sets of functions
\begin{equation*}
   \{\mu_n^c;\quad c=1,2,3,...,k\},
\end{equation*}
where
\begin{equation*}
    \mu_n^c = \sum_{i=1}^nX_i(t)\textbf{1}_{\{C_i=c\}}/N_c ,
\end{equation*}
and 
\begin{equation*}
    N_c = \sum_{i=1}^{n}\textbf{1}_{\{C_i = c\}}.
\end{equation*}

\par
\begin{figure}[!t]
\centerline{\includegraphics[width=\columnwidth]{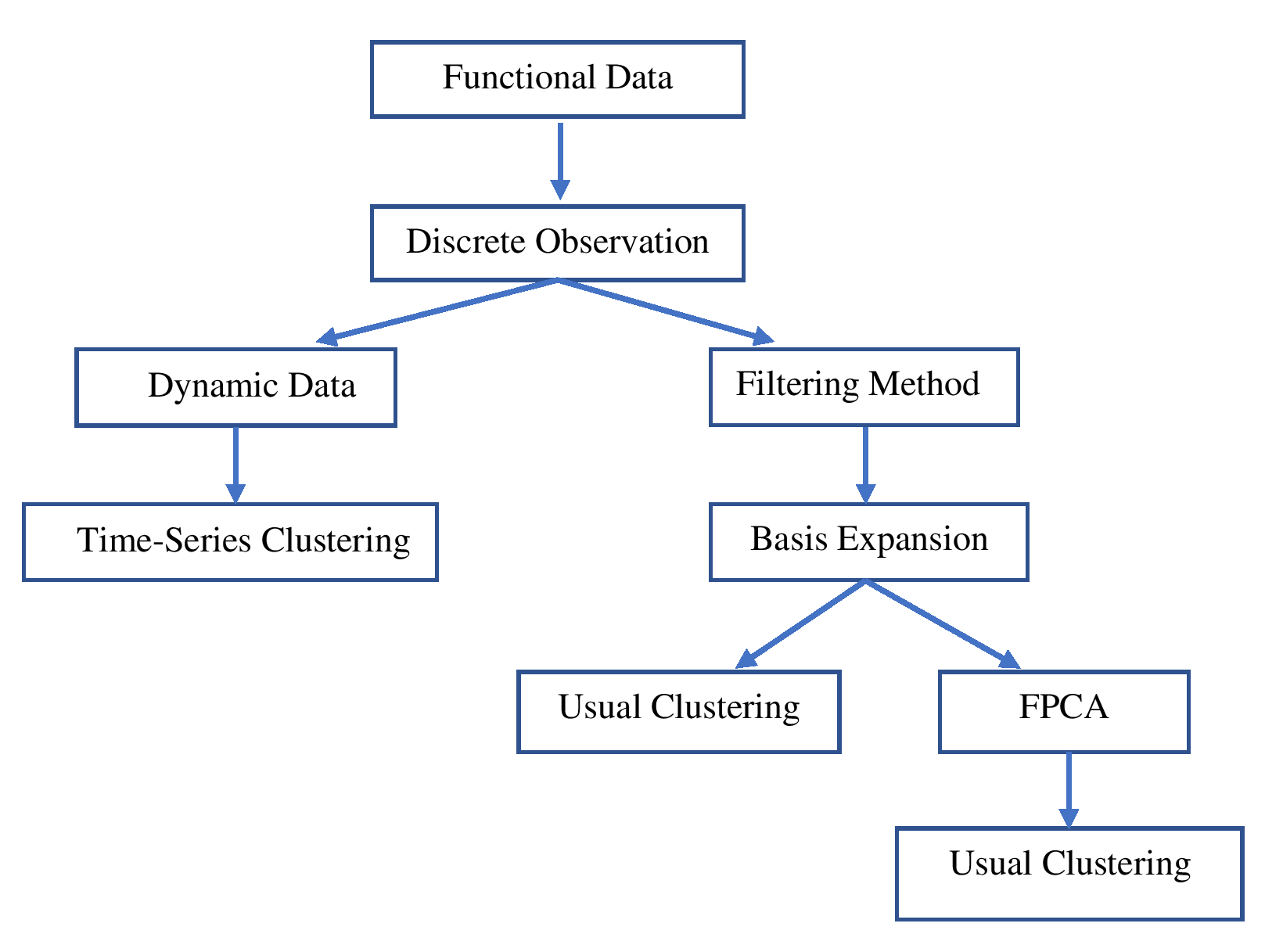}}
\caption{Clustering Scheme}
\label{Clustering Scheme}
\end{figure}

 In \cite{b10},\cite{b14}, and \cite{b15} the authors have given a broad idea for clustering approach to the functional data and also discussed about the advantages and limitations of different approaches. Fig. \ref{Clustering Scheme} outlines our scheme to cluster the data set associated with patients facial paresis.

%%%%%%%%%%%%%%%%%%%%%%%%%%%%%%%%%%%%%
\subsubsection{Time-Series clustering} \label{TS clustering}
Time-series data is a sort of dynamic data that typically results from the continuous recording of events, such as medical data, economic data, and market data, to name a few examples. Clustering algorithms for dynamic data require extra care due to its high dimensionality, where dimensionality is defined as the length of the series. In most cases, the functional data are collected at several discrete values in a domain. Under this approach, we do not consider the functional nature of the data but rather treat it as temporal data and clustering approach for time series are utilized to cluster the data.  
\par Time series data clustering necessitates dissimilarity measures; the most popular approaches are hierarchical, partitional, and fuzzy clustering. For the clustering of time series, we have use the tsclust function of dtwclust package in the statistical software R \cite{b16}. Traditionally, clustering algorithms assign one observation to only one cluster, a process known as hard clustering. Another approach is to do fuzzy partitioning of the observation, which involves assigning the item to more than one cluster at a time with a particular membership value (the sum of the membership value is not greater than one), called fuzzy clustering. We have used Fclust function of fclust package in statistical software R to perform fuzzy clustering of our data \cite{b17}. 
%%%%%%%%%%%%%%%%%%%%%%%%%%%%%%%%%%%%%%%%%%

\subsubsection{Filtering method} \label{Filtering method}
In the filtering method, we aim to reduce the dimension of the data by expressing the data into finite set of basis functions such as B-spline. Another approach is reconstruct the data into its functional nature using a set of basis functions. Then, considering the finite number of functional principal components, we can apply the usual clustering methods. The step that involves dimension reduction of the functional data is called the filtering step for the clustering method. The following two filtering methods are used in the present work \cite{b18,b19}.    
\begin{itemize}
    \item  \textbf{Clustering via Basis expansion:} As discussed in \ref{FDA}, the set $\beta = \{\beta_1, \beta_2,...,\beta_k\}$ represents a basis function of the finite-dimensional space. We can approximate the curve by equation \ref{eq:1}, where we approximate the curve by the first $k$ projection of the curve onto the space spanned by $\beta$. We apply the clustering algorithms to the set of coefficients of the curves in the basis representation. We have used the funFEM function of the funFEM package in the statistical software R \cite{b20}.
    \item  \textbf{Clustering via FPCA:} Followed by the basis representation of the functional data, we apply the popular dimension reduction method for functional data, i.e., functional principal component analysis (FPCA). The FPCA were calculated using fda package in the statistical software R \cite{b21}. We will utilize the first $p$ principal components for the usual clustering algorithms, such as $k$-means, Hierarchical clustering, to find the clusters of the functional data. For the filtering method with FPCA, we have applied Hierarchical and $k$-means clustering algorithms from the stat package, the Partitioning Around Medoids (PAM) function from the cluster package, and  the Mclust function from the mclust package in the statistical software R \cite{b22,b23}.
\end{itemize}

\section{Result} \label{result}
\begin{figure}[!t]
\centerline{\includegraphics[width=\columnwidth]{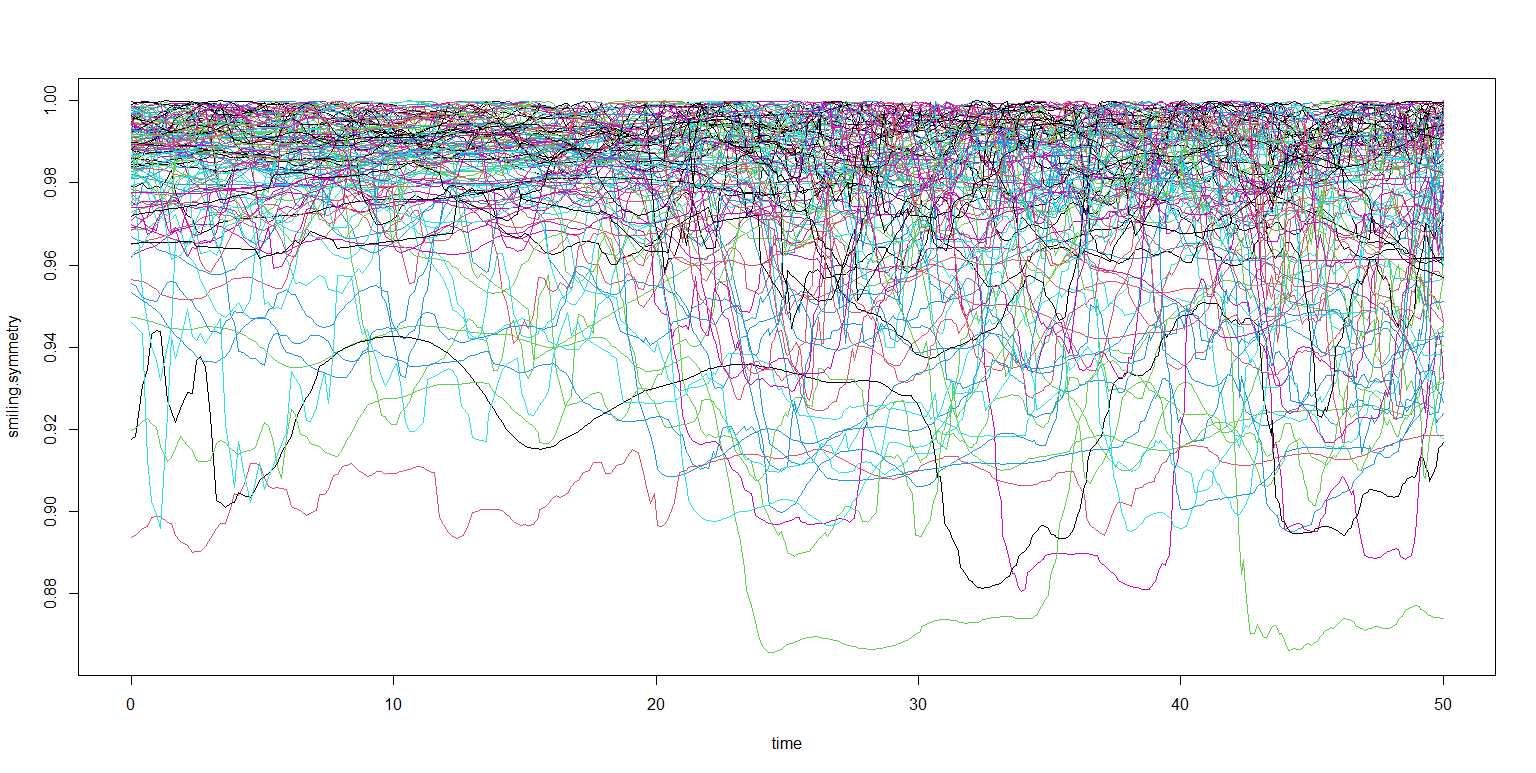}}
\caption{Indicator: smiling.symmetry of all 120 patients.}
\label{fig:smiling symmetry}
\end{figure}

As discussed in \ref{method}, smiling.symmetry is the indicator which has maximum Spearman's correlation value with HB grade. Therefore, in the present work we cluster the patients based on only one indicator curve i.e., smiling.symmetry. We make 4 clusters of the patients based on increasing degree of facial paresis and compare the outcome with the adjusted clinical label data (Table. \ref{Tab5}).  Fig. \ref{fig:smiling symmetry} shows the curve representation of  smiling.symmetry of all 120 patients.

\vspace{1cm}
\textbf{Method 1:}
\par
\begin{figure}[!t]
\centerline{\includegraphics[width=\columnwidth]{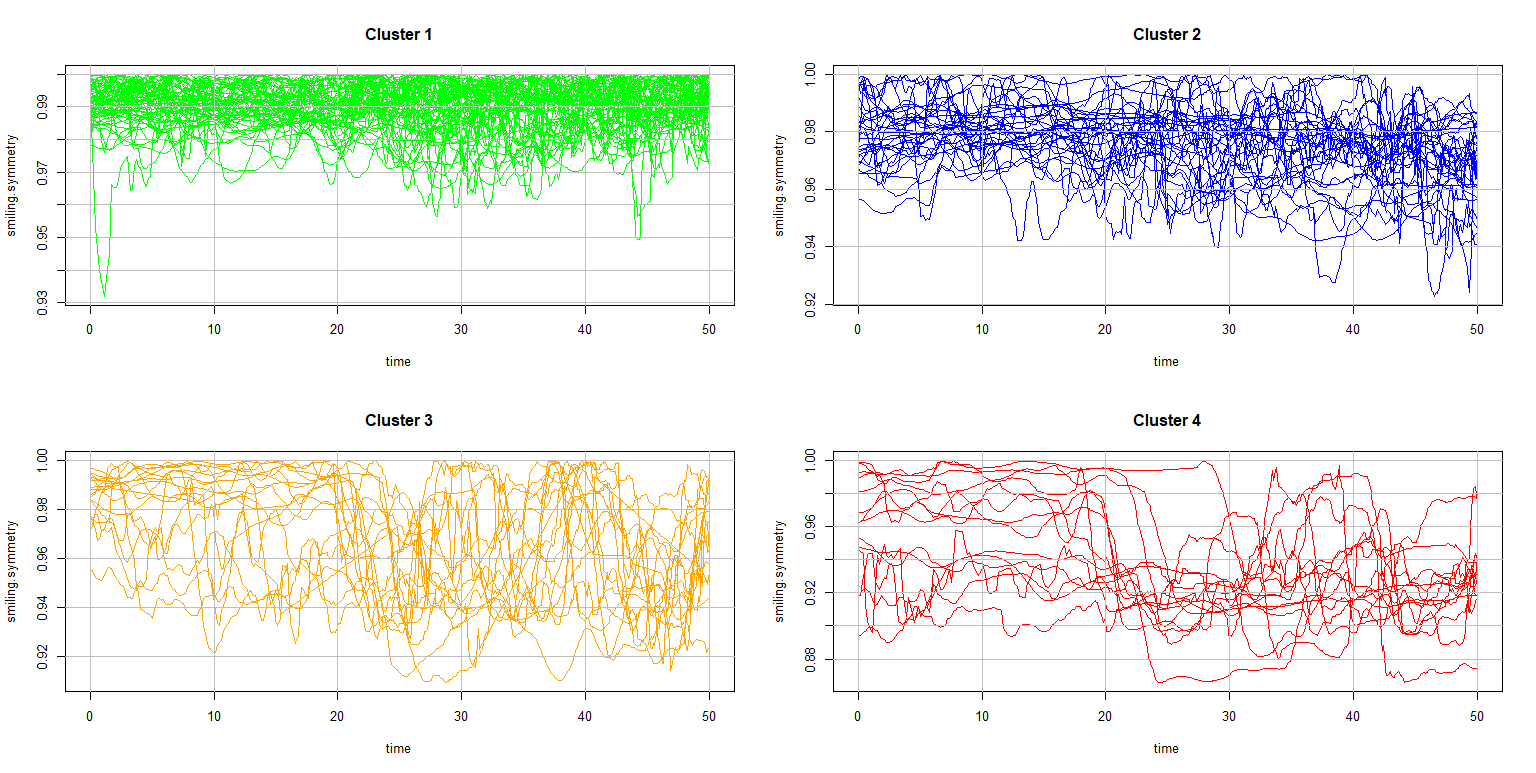}}
\caption{Four clusters of the patients based on smiling.symmetry by TS clustering.}
\label{fig:cluster_1}
\end{figure}

 The clustering of patients based on smiling.symmetric is performed by dynamic data clustering method of the time series using dynamic time warping, and found  4 clusters with 57, 32, 17 and 14 members and these cluster's are shown in Fig. \ref{fig:cluster_1}.
\par  It can be observed that first cluster has most number of patients and the symmetry value are closer to one i.e, high symmetry when performing exercise smiling,  we can say that this group has better symmetry and it represents the patients with HB1 grade. Cluster 2 have second most number of patients with symmetry value slightly less than one, therefore this cluster represents patients with HB2 grade. Similarly cluster 3 and cluster 4 represents patients with  HB3 and HB6 grades respectively, due to the fact that the symmetry value are much less when compare to cluster 1 and cluster 2. Table \ref{Tab6} shows the contingency table for the above. 

\begin{table}
\centering
\caption{ Contingency tables of TS clustering.}
\begin{tabular}{p{1.5cm} p{1.5cm}p{1.cm}p{1.cm}p{1.cm}p{1.cm}}
& & \multicolumn{4}{p{110pt}}{\centering Label}  \\
\hline
 &  & HB1& HB2 &HB3&HB4\\
 \hline 
\multirow{4}{*}{Clusters} & 1 & 33 & 16 & 8 & 0 \\
& 2 & 21 & 3 & 6 & 2 \\ 
& 3 & 2 & 1 & 6 & 8\\
& 4 & 1& 0 & 3 &10\\
 \hline
\end{tabular}
\label{Tab6}
\end{table}
%%%%%%%%%%%%%%%%%%%%%%%%%%%%%%%%%%%%%%%%%%%%%%%%%%%%%%%%%%%%
\vspace{1cm}
\textbf{Method 2:}
\par Another approach under the Time-series clustering method is fuzzy $k$-means (FKM). Instead of assigning patients to a fixed cluster, FKM assigns a function to more than one cluster with a membership score ranging from 0 to 1 that indicates how strongly the patients  belong to a specific cluster.
\par
\begin{figure}[!t]
\centerline{\includegraphics[width=\columnwidth]{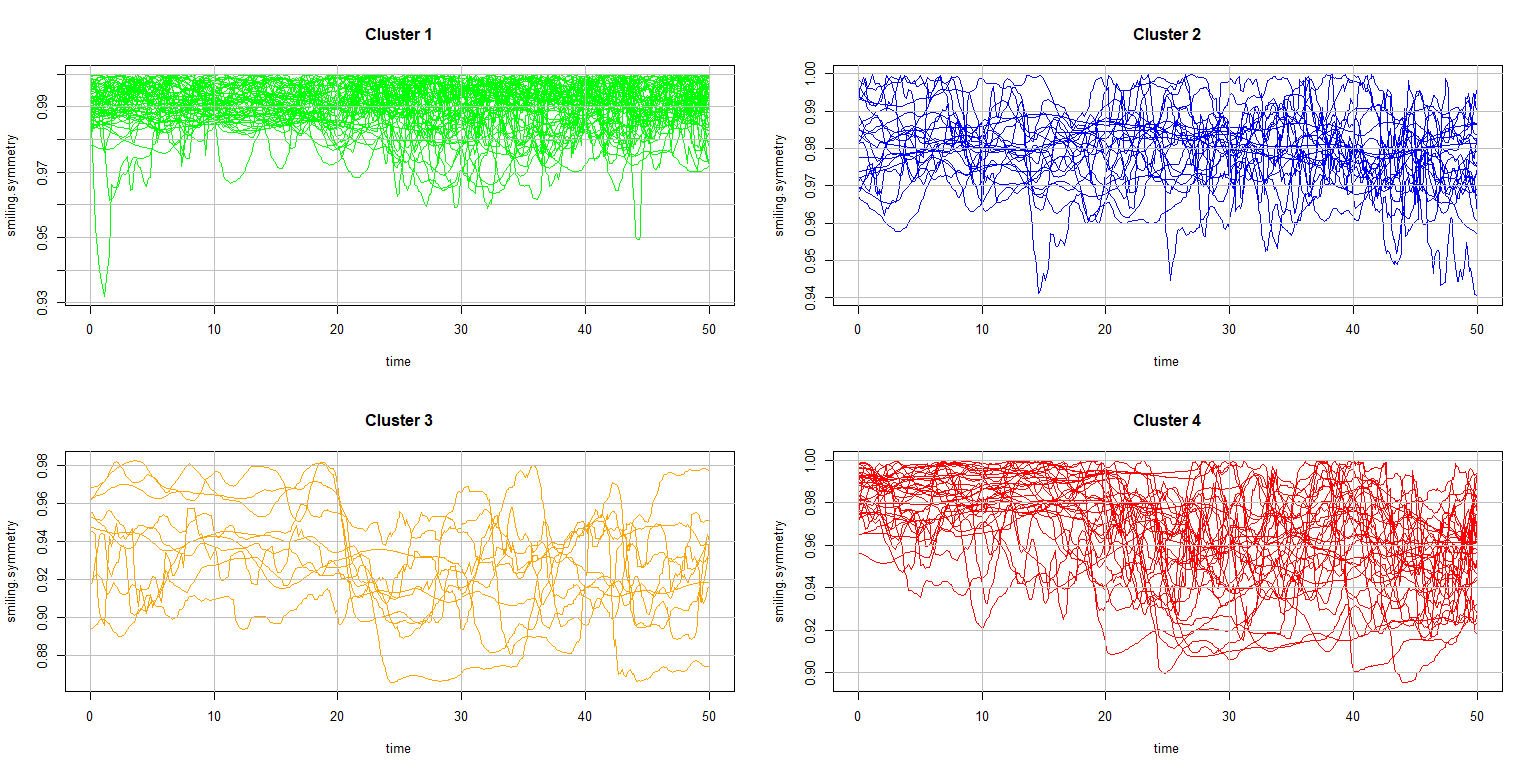}}
\caption{Four clusters of the patients based on  smiling.symmetry by FKM.}
\label{fig:cluster_2}
\end{figure}

\begin{figure}[!t]
\centerline{\includegraphics[width=\columnwidth]{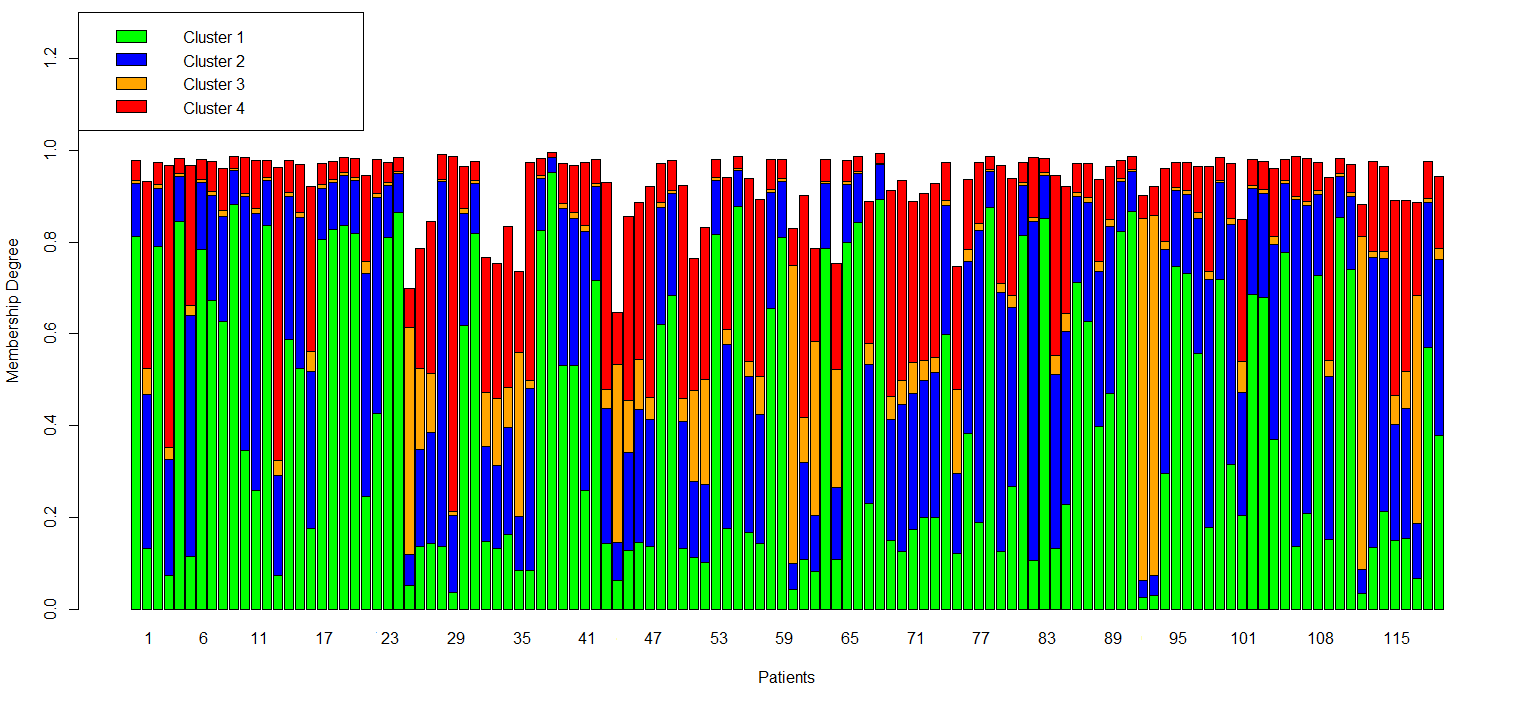}}
\caption{Membership Degree plot of FKM clustering.}
\label{fig:Membership plot}
\end{figure}

 FKM divides the patients into four clusters with 55, 22 ,10, and 33 members. The clusters are shown in  Fig. \ref{fig:cluster_2}. Cluster 1 represents patients from the HB1 class due to the high symmetry value for the smiling exercise, and similarly, cluster 2, cluster 3, and cluster 4 represent patients from HB2, HB3, and HB6 classes, respectively. The contingency table for the same is shown in Table. \ref{Tab7}. Since FKM allows patients to more than one cluster at a time, we do have a membership degree value for each patient for the clusters, and Fig. \ref{fig:Membership plot} represents the membership degree of each patient belonging to cluster 1, cluster 2, cluster 3, and cluster 4.     

\begin{table}
\centering
\caption{ Contingency tables for the FKM.}
\begin{tabular}{p{1.5cm} p{1.5cm}p{1.cm}p{1.cm}p{1.cm}p{1.cm}}
& & \multicolumn{4}{p{110pt}}{\centering Label}  \\
\hline
 &  & HB1& HB2 &HB3&HB4\\
 \hline
\multirow{4}{*}{Cluster} & 1 & 31 & 16 & 8 & 0 \\
& 2 & 17 & 0 & 4 & 1 \\ 
&3 & 2 & 0 & 2 & 6\\
&4 & 7& 4 & 9 &13\\
 \hline
\end{tabular}
\label{Tab7}
\end{table}

%%%%%%%%%%%%%%%%%%%%%%%%%%%%%%%%%%%%%%%%%%%%%%%%%%%%%%%%%%%%%%%
\vspace{1cm}
\textbf{Method 3:}
\par
\begin{figure}[!t]
\centerline{\includegraphics[width=\columnwidth]{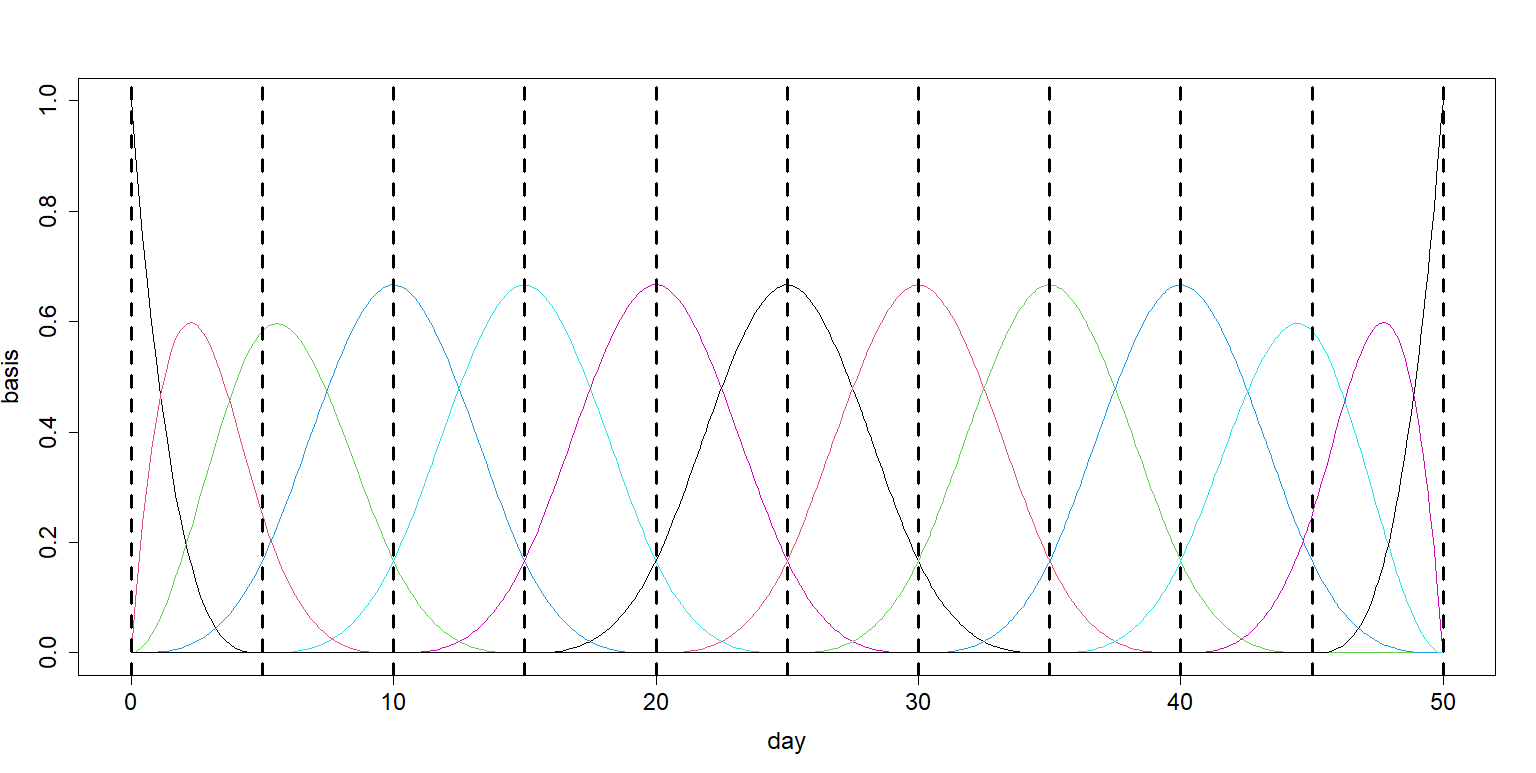}}
\caption{Basis functions of order 4 spline and 9 interior knots (vertical dashes line).}
\label{fig:B-spline basis}
\end{figure}

 As discussed in section \ref{FDA}, the functional data need to be seen as a single object instead of a sequence of observations. Therefore we represent the discretely observed smiling.symmetry data of patients using the basis function and expressing them in functional form. For this purpose, we use the B-spline basis function shown in Fig. \ref{fig:B-spline basis}. 
\par
\begin{figure}[!t]
\centerline{\includegraphics[width=\columnwidth]{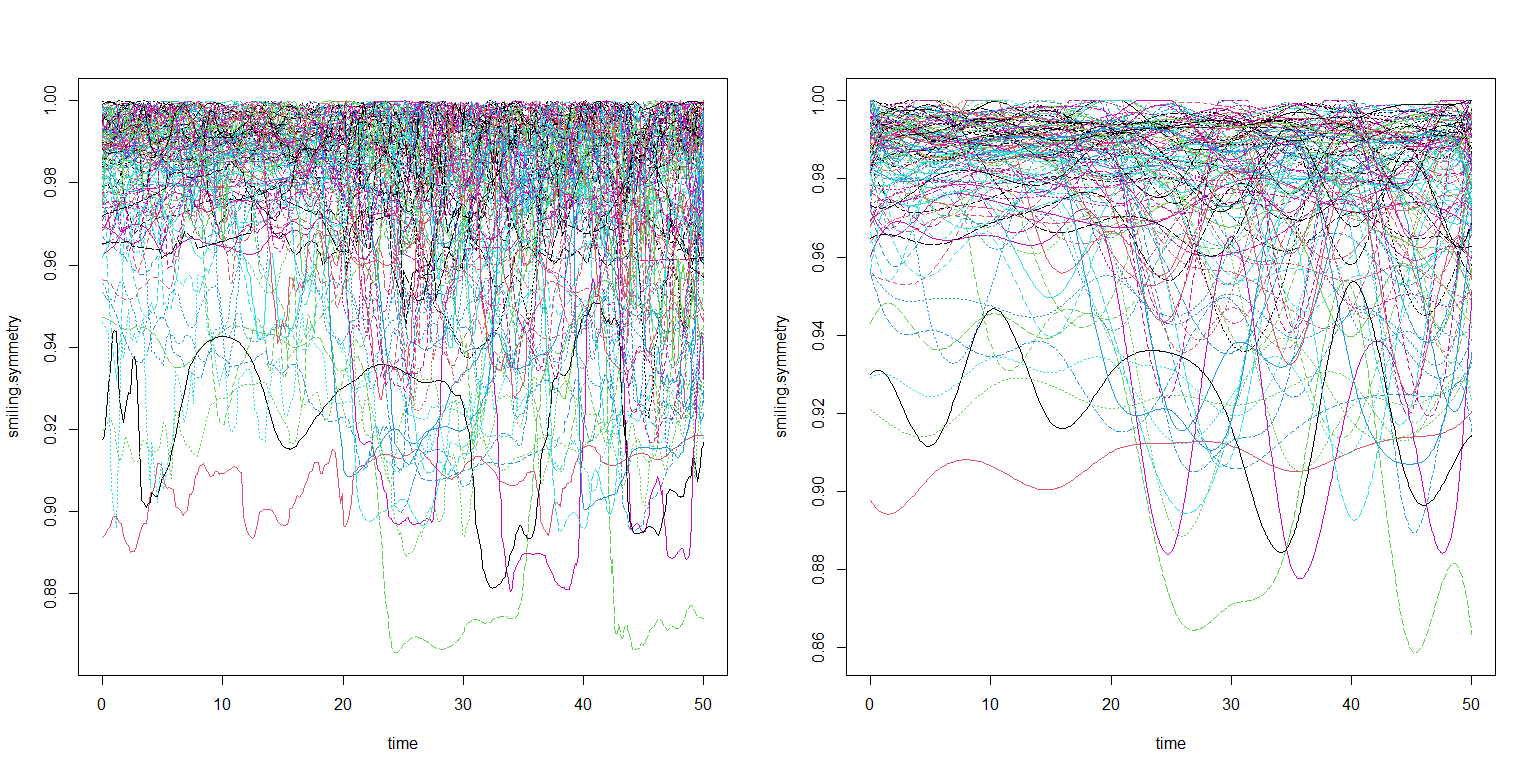}}
\caption{smiling.symmetry in discrete form (left) and functional form (right).}
\label{fig:Smooth functional data}
\end{figure}

 Expressing the functional data using the B-spline, the curve now be represented by 13 coefficients reduce the dimension of the functional data significantly and provide smooth functions (see Fig. \ref{fig:Smooth functional data}).
\par After the filtering process, i.e., converting the data into its functional form using basis functions, we apply the clustering method for functional data.

\par
\begin{figure}[!t]
\centerline{\includegraphics[width=\columnwidth]{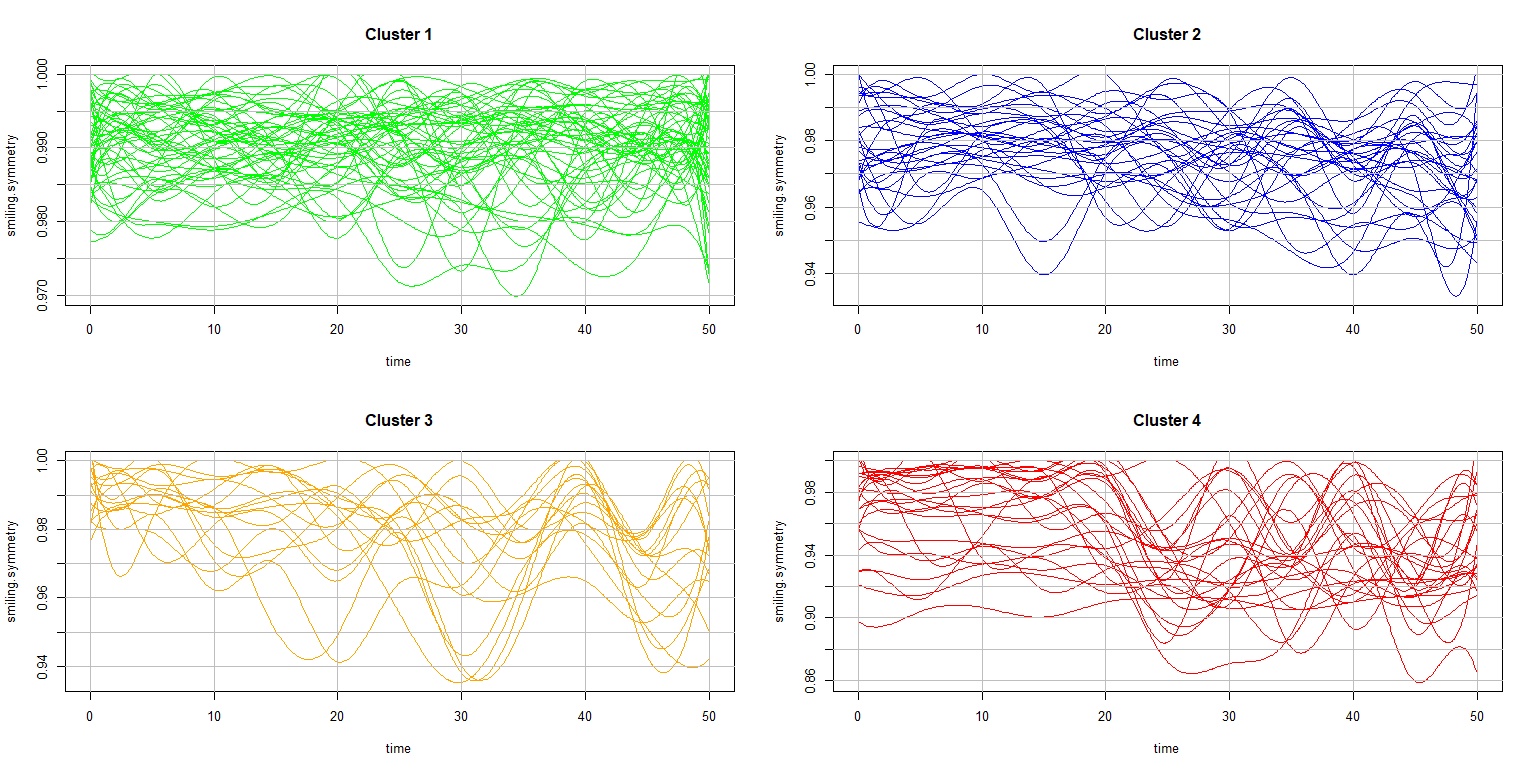}}
\caption{Four clusters by filtering method.}
\label{fig: Cluster_3}
\end{figure}

 The algorithm divides the functional data into 4 clusters with 50, 29, 16, and 25 members (See Table. \ref{Tab8}). We can visualize the clusters in Fig. \ref{fig: Cluster_3}. The value of symmetry.smiling for cluster 1 is close to one compared to the values for clusters 2, 3, and 4. Since the group with the highest smiling.symmetry must have a lower HB grade compared to the groups with the lower value of smiling.symmetry. Therefore, we may conclude that cluster 1 represents patients from HB1 grade, and cluster 2, cluster 3, and cluster 4 represent patients from HB2, HB3, and HB6 grades respectively.

\begin{table}
\centering
\caption{Contingency tables for the filtering (basis) methods.}
\begin{tabular}{p{1.5cm} p{1.5cm}p{1.cm}p{1.cm}p{1.cm}p{1.cm}}
& & \multicolumn{4}{p{110pt}}{\centering Label}  \\
\hline
 &  & HB1& HB2 &HB3&HB4\\
 \hline
\multirow{4}{*}{Cluster}  & 1 & 31 & 15 & 4 & 0 \\
& 2 & 18 & 3 & 6 & 2 \\ 
&3 & 5 & 2 & 7 & 2\\
&4 & 3& 0 & 6 &16\\
 \hline
\end{tabular}
\label{Tab8}
\end{table}

%%%%%%%%%%%%%%%%%%%%%%%%%%%%%%%%%%%%%%%%%%%%%%%%%%%%%%%%%%%%
\vspace{1cm}
\textbf{Method 4:}
\par
\begin{figure}[!t]
\centerline{\includegraphics[width=\columnwidth]{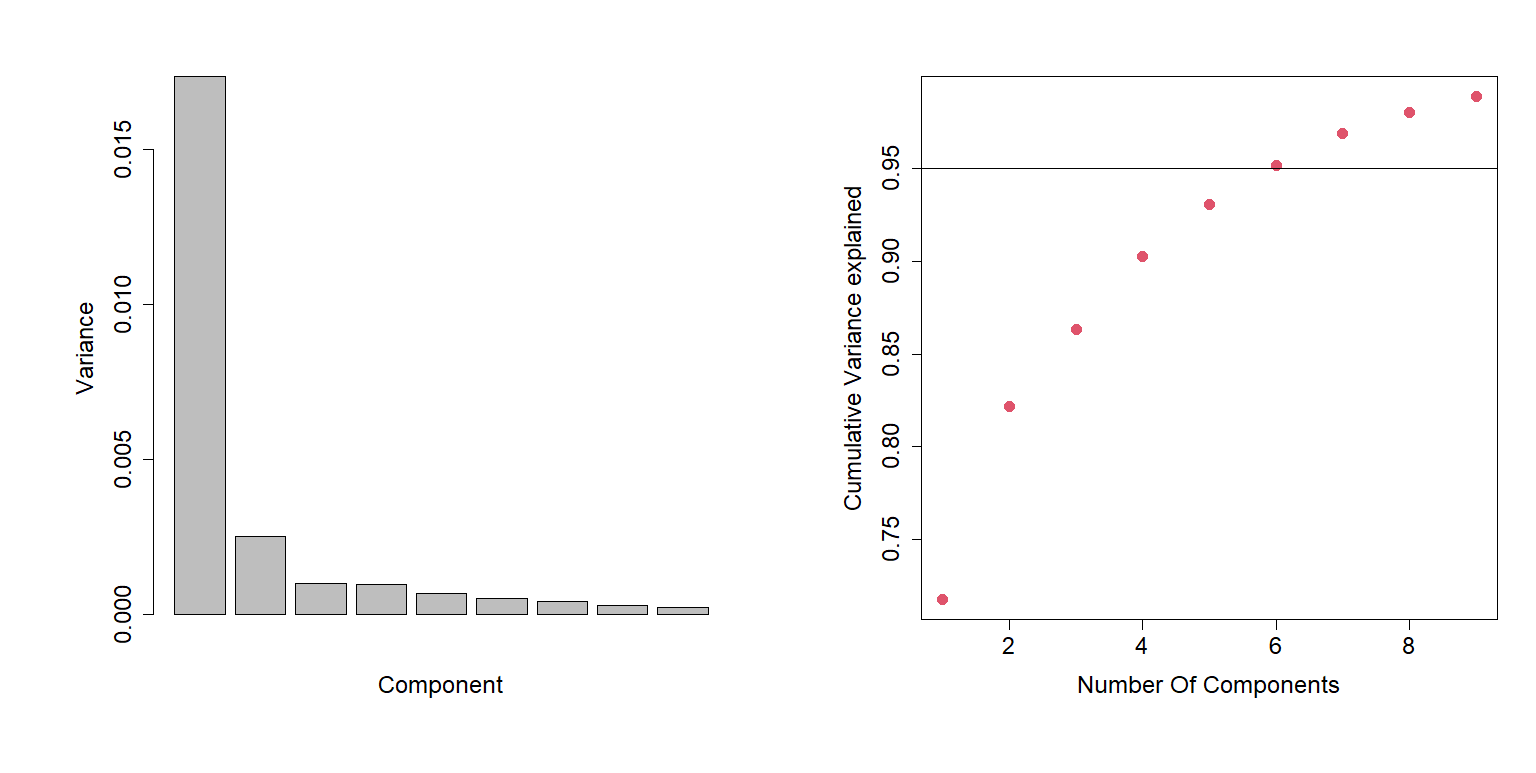}}
\caption{Variance (left) and Cumulative variance (Right) of principal components.}
\label{fig:variance of eigen-vector}
\end{figure}

\begin{figure}[!t]
\centerline{\includegraphics[width=\columnwidth]{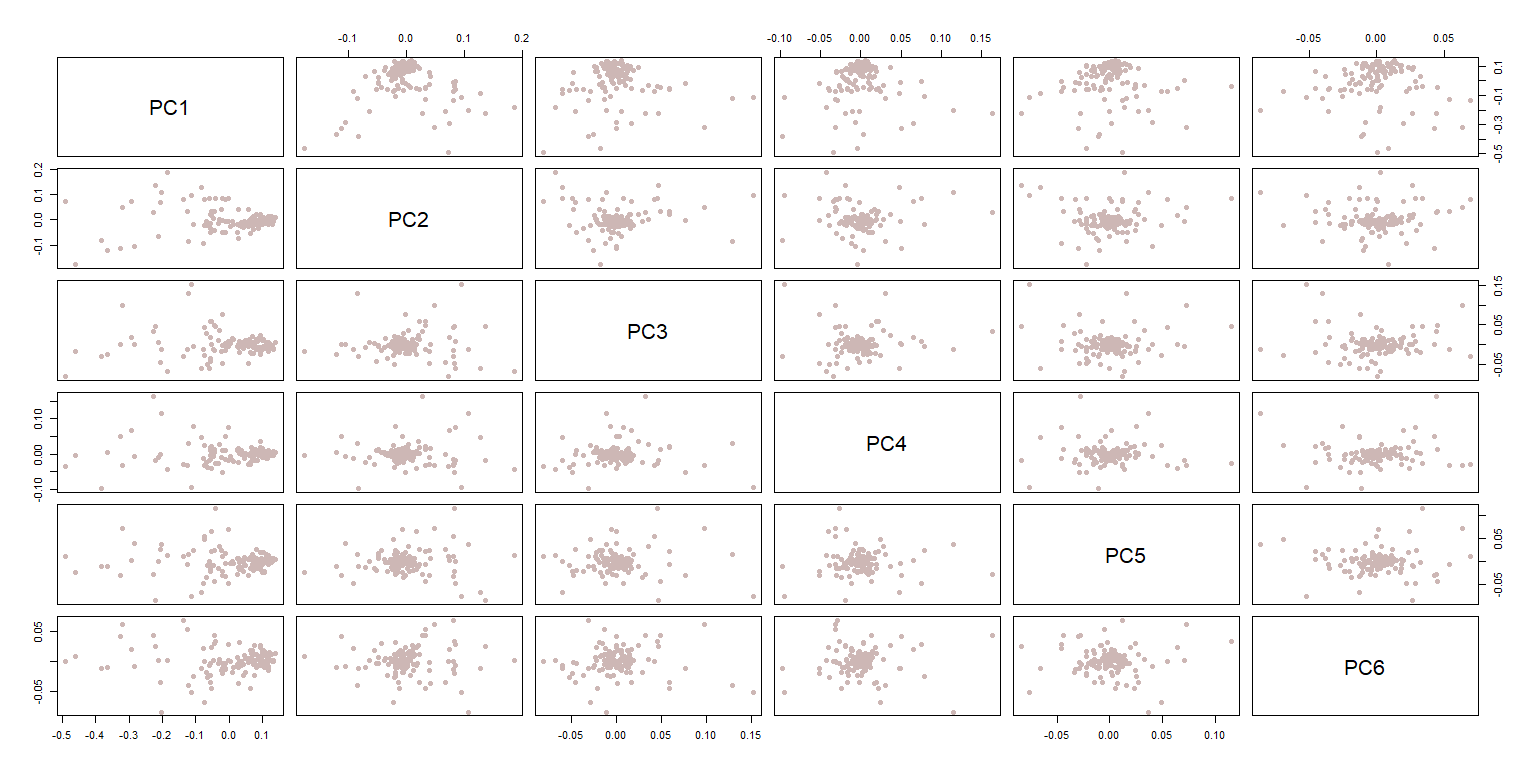}}
\caption{scatterplots of projected data in the 6 first principal components.}
\label{fig:All PCs}
\end{figure}

 Following the representation of data using B-spline, we find the Functional principal components (FPCs) of the functional data smiling.symmetry and then apply the clustering algorithm. The FPCs summarise the functional and simultaneously reduce the dimension of the functional data. Fig. \ref{fig:variance of eigen-vector} shows that only six principal components explain more than $95\%$ of the cumulative variance. Fig. \ref{fig:All PCs} shows the scatter plot of all six PCs of the reduced functional data.
\par We projected the data onto the first six principal components (having more than $95\%$ variance), i.e., reduced the functional data into finite-dimensional data and applied usual clustering algorithms such as Hierarchical, $k$-means, Partitioning Around Medoids (PAM), and Model based clustering (Mclust).
\par
\begin{figure}[!t]
\centerline{\includegraphics[width=\columnwidth]{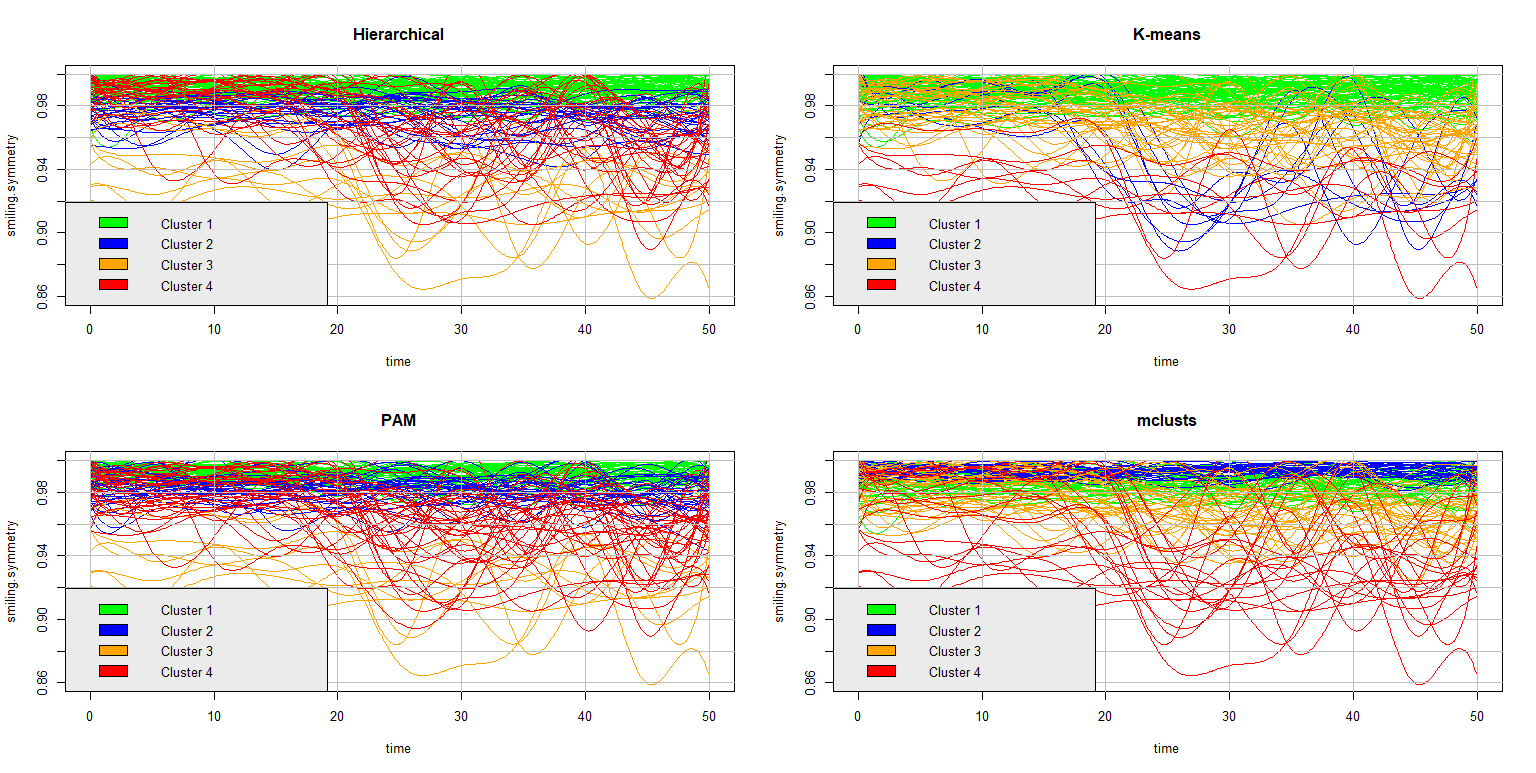}}
\caption{Cluster plot of all FPCA methods.}
\label{fig:cluster_4}
\end{figure}

 Hierarchical clustering divides patients into four clusters with 57, 24, 10, and 29 members. Similarly, k-means divides the patients into 69, 8, 35, and 8 members, PAM divides the patients into 47, 29, 9, and 35 members, and Mclust divides the patients into 42, 25, 32, and 21 members. The clusters are shown in Fig. \ref{fig:cluster_4}. See Table. \ref{Tab9} for the details of the result of all the clustering algorithms.

\begin{table}
\centering
\caption{Contingency tables for the filtering (FPC) methods.}
\begin{tabular}{p{1.5cm} p{1.5cm}p{1.cm}p{1.cm}p{1.cm}p{1.cm}|p{1.cm}p{1.cm}p{1.cm}p{1.cm}}
& & \multicolumn{4}{p{70pt}}{\centering Label}& \multicolumn{4}{p{70pt}}{\centering Label} \\
 \hline
 &  &\multicolumn{4}{c}{(a) Hierarchical} &\multicolumn{4}{c}{(b) $k$-Means}  \\ 
\hline
 &  & HB1& HB2 &HB3&HB4& HB1& HB2 &HB3&HB4\\
 \hline
\multirow{4}{*}{Cluster} & 1&32&16&9&0&43&16&10&0\\ 
&2&17&2&3&2&1&0&2&5\\
&3&2&0&2&6&12&4&10&9\\
&4&6&2&9&12&1&0&1&6\\
 \hline
  &  &\multicolumn{4}{c}{(c) PAM}  &\multicolumn{4}{c}{(d) Mclust}  \\ 
 \hline
\multirow{4}{*}{Cluster} &1&28&15&4&0&27&8&7&0\\
&2&19&1&8&1&15&7&3&0\\
&3&2&0&1&6&13&5&7&7\\
&4&8&4&10&13&2&0&6&13\\
 \hline

\end{tabular}
\label{Tab9}
\end{table}

\section{Discussion} \label{Discussion}
We have discussed different unsupervised methods to distinguish the patients based on the mimetic movement of patients post-head and neck surgery. HB grade is the most widely used scale by the clinician to categorize facial nerve dysfunction, but with a high degree of subjectivity. This work aims to develop a method that can perform this task more objectively and ultimately save the clinician's time. 
\par We performed an independent analysis from clinician labels of the rehabilitation process only based on one indicator, ``smiling.symmetry" and compared it with the clinician analysis. We perform the cluster analysis for a different form of functional data, such as: 
\begin{itemize}
    \item Time-Series clustering
    \item Basis expansion
    \item Clustering based on functional principal component analysis
\end{itemize}
The results are not the same for all the clustering algorithms used for different forms of functional data like dynamic form, basis representation, or when reduced to finite-dimensional data by using FPCA. On comparing the results with the HB grades, it can be seen that all algorithms for different data form distinguish the patients from  HB1 and HB6 grades quite well, but faces some difficulties in grading the patients who had adjacent HB grade. Table. \ref{Tab10} presents a summary of all the clustering methods used in this work.
\par We employ the correct classification rate (CCR) to gauge the appropriateness of the generated clusters with the available partition. Since, our major objective is not to predict the HB Grades because it could be subjective, we consider the correct classification as well as approximate classification of patients. We define approximate CCR in which members of the nearest clusters are also considered in the cluster. For example, for cluster 1, both cluster 1 and cluster 2 are considered, and for cluster 3, we consider cluster 2, cluster 3, and cluster 4 as one cluster. For all the clustering methods, silhouette width is also calculated.
\par On comparing the result with previous work of \cite{b1}, which uses 14 indicators  (cf. Table 8 in \cite{b1}) to do the classification analysis and obtained accuracy of $60\%$ and approximate accuracy of $86\%$, we uses only one indicator and got almost $45\%-50\%$ accuracy for some of the methods but better approximate CCR of $89.4\%$ with funFEM and $87.42\%$ with dtwclust.

\begin{table}
\centering
\caption{ Correct classification rates (CCR), Approximate CCR in percentage and Silhoutte-width of all the clustering algorithm used in this paper.}
\begin{tabular}{p{4cm}p{3cm}p{3.5cm}p{3cm}}
 \hline
 Methods& CCR (in \%) & Approximate CCR (in \%)& Silhoutte width  \\ 
\hline
\hline
Time-Series clustering && &\\
\hline
dtwclust& 0.4333& 0.8742&0.41\\
Fclust& 0.3833&0.8105&0.35\\
\hline
Filtering (Basis)&  & &\\
\hline
funFEM &0.475& 0.894&0.27\\
\hline
Filtering (FPCs)& & &\\
\hline
Hierarchical&0.4&0.8567&0.45\\
k-means&0.4917&0.7717&0.52\\
pam & 0.3583& 0.85&0.32 \\
Mclust& 0.45&0.8567&0.2\\
\hline
\hline
\end{tabular}
\label{Tab10}
\end{table}

\section{Future Work} \label{future work}
So far in this paper we used the information  of  mimetic muscles while performing a certain set of given  exercises. The indicator considered in this work  is smiling.symmetry because of the maximum Spearman's correlation with the HB grade of the patients. Although we have certain more indicator of patients which can also gave better method of clustering but that will make the data set high dimensional and which requires advanced functional technique and analysis method. 
\par In future we will utilize all the given indicators of a patients such as teeth.symmetry, lips.symmetry, frowning.symmetry and eyebrow.symmetry. Using all indicators at same results into multivariate functional data and may give better result compared to the present results.

\vspace{1.5cm}

\textbf{Acknowledgements}
\par This work is supported by the Department of Science and Technology, Govt of India under the project Grant No. DST/INT/CZECH/P-10/2019 (Indo-Czech Bilateral Research Program). 

%%%%%%%%%%%%%%%%%%%%%%%%%%%%%%
%%%%%%%% Bibliography 

\end{document}